\documentclass[runningheads]{llncs}
\usepackage[T1]{fontenc}
\usepackage{graphicx}
\usepackage{array}
\usepackage{enumitem}
\usepackage{graphicx}
\usepackage{booktabs}
\usepackage{subcaption}
\usepackage{comment}
\usepackage{multirow}
\usepackage{amssymb}
\usepackage{tikz}
\usepackage{orcidlink}
 
\usepackage{bbding}

\newcolumntype{L}[1]{>{\raggedright\arraybackslash}p{#1}}

\newlist{dashlist}{itemize}{1}
\setlist[dashlist]{
    label={-},
}

\newcommand{\circled}[1]{%
  \tikz[baseline=(char.base)]\node[shape=circle,draw,inner sep=1pt] (char) {#1};}

\usepackage{url}
\begin{document}
\title{Developing a Roadmap to an AI-first Organization: A Case Study in Embedded Software Development}
\titlerunning{Roadmap to an AI-first Organization}
%
\author{Viktor Kjellberg\orcidID{0009-0006-4330-2385}\Envelope \and
Srijita Basu\orcidID{0000-0002-6835-947X} \and
Simin Sun\orcidID{0009-0003-0873-5968} \and
Farnaz Fotrousi\orcidID{0000-0001-5385-0381} \and
Miroslaw Staron\orcidID{0000-0002-9052-0864}
}

\authorrunning{V. Kjellberg et al.}
%
\institute{University of Gothenburg and Chalmers University of Technology
\email{\{viktor.kjellberg,srijita.basu,simin.sun,farnaz.fotrousi,miroslaw.staron\}@gu.se}}

\maketitle              
\begin{abstract}

The emergence of AI agents is expected to reshape software engineering by moving beyond AI as assistants towards systems capable of planning, executing, and evaluating development tasks with increasing autonomy. This transition is particularly significant for embedded software organizations, where strict requirements for quality, traceability, verification, and long-term maintainability often apply. This paper presents a case study of a large embedded systems company and its transition toward becoming an AI-first organization. Through a mixed method, we analyzed data collected from a semi-structured workshop with 40 participants, including scrum masters, architects, management, and product owners. The findings show that the participants expect agentic AI to affect team structure, required competencies, organizational strategies, and developers' roles within the organization. Based on these findings, the paper discusses implications for federated AI team formation, human-in-the-loop practices in such an organization, and the sustainable adoption of AI agents in embedded software engineering. We also present a concrete roadmap for the organization towards becoming an AI-first organization.

\end{abstract}
\section{Introduction}

AI-based systems support developers in everyday tasks such as code generation, debugging, and documentation. In particular, tools such as GitHub Copilot have transformed developers' work practices by generating code from natural-language prompts and contextual input \cite{10822885}. These systems can accelerate software development by enabling closer collaboration between developers and AI systems \cite{10.1145/3652154}. So far, the use of such tools has been user-centric, emphasizing the role of the human-in-the-loop and positioning Large Language Models (LLMs) as technologies that augment, rather than replace, software engineers \cite{10.1145/3652154}.

With the emergence of AI agents, software companies enter another transformative period. Unlike interactive AI tools that primarily support developers through collaboration, AI agents promise greater, long-horizon autonomy, including the ability to plan, execute, and evaluate tasks with limited human intervention \cite{wang2026doesagentdevelopmentreflect}. However, the extent to which such agents can be integrated into existing software engineering workflows depends on several contextual factors: company's AI-readiness, tools support, preparation of the source code-base.

In embedded systems development, LLM-based systems have shown promise for supporting both code generation and debugging, even when the generated code contains errors \cite{10.1145/3613905.3650764}.

Moreover, AI agents can mitigate some of the drawbacks of LLMs by combining external tools to validate and test the generated code \cite{11344427}. At the same time, the introduction of agents into embedded systems development requires careful consideration, as explained by Sun and Staron \cite{sun2026agenticpipelinesembeddedsoftware}. To take full advantage of these systems' capacity, a few emerging protocols are changing how software should be developed. AI-friendly Code Protocols (AICP) govern how data should be represented and structured to ease integration with Model Context Protocol (MCP) and similar infrastructures. AI-friendly Code also refers to the structure of the code itself, ensuring it is context-rich with embedded metadata for LMMs to extract and use as context. 
However, companies can take different approaches to how this should proceed. The approach can be either top-down, driven by centralized leadership, or bottom-up, driven by experimentation by the practitioners. Both have clear advantages and disadvantages.  
Sun and Staron \cite{sun2026agenticpipelinesembeddedsoftware} offer insight into how some companies developed their strategy for this shift, but raise the question of how companies view their transformation to the next step in the landscape of software evolution and how far they want to go in this transformation.

This transition towards what is referred to as Software 4.0 (SE 4.0) \cite{11319541} presents both benefits and challenges for a company beginning its journey. One of these is the need to redesign processes with an AI-first focus, meaning the processes should be designed to optimize the performance of AI agents \cite{TowardsAIOrganizations}. In this paper, we present the views and vision of a large embedded systems company on how Agentic systems are envisioned to affect their processes and workflows over the next five years. We discuss the adaptations they envision and their journey toward becoming an AI-first organization. All towards answering the following research question:

\begin{itemize}

    \item RQ1: How do embedded software development engineers envision the transition to an AI-first organization?

\end{itemize}

To answer this overarching question, we address the following sub-questions:

\begin{itemize}

    \item RQ1.1: Which technical, organizational, and human factors are expected to be affected by the transition?

    \item RQ1.2: What challenges should be considered when developing the roadmap for such a transition?

\end{itemize}

This paper makes three contributions. First, it reports empirical evidence from an industrial case study on practitioners' perceptions of how AI agents will affect their organizations in the coming years. Second, we characterize the main adoption challenges and opportunities that emerged from a structured workshop. Lastly, we synthesize these findings into a practitioner-oriented roadmap that outlines how the organization can transform.

\section{Background: Evolution of Software}
Since the introduction of transformer-based LLMs \cite{NIPS2017_3f5ee243}, the workflows in Software Engineering (SE) have changed rapidly. Many software companies have already moved to what Jiang et al. \cite{11319541}, called SE 3.0, where LLMs, through prompting, are used to support a variety of SE tasks, everything from code completion and code reviews to PR summaries, to increase the productivity of developers \cite{sergeyuk_using_2025,10.1145/3796563.3796601}. Alongside increasing development speed, LLM-based assistant tools enable learning opportunities through providing new alternative solutions for developers \cite{ferino2025walkingtightropellmssoftware,10.1145/3643795.3648377}. The initial versions of GitHub Copilot can be seen as an example of a tool that fits into SE 3.0, where it is integrated into production but remains fully reliant on human input.

With the introduction of frameworks that enable LLMs to connect to tools, such as MCP, which allows the creation of AI agents, we entered a new phase, SE 4.0 \cite{11319541}. Here, the AI agents are granted greater autonomy to interact with external tools without a developer as a middleman. However, AI agents are still limited in the number of tasks and tools they can use. They still require user input on which tasks they are supposed to solve. So, in other words, they are allowed greater autonomy in how they solve tasks, but not in what they are supposed to solve. For some, the shift is limited to low-risk activities, such as documentation generation, pull request analysis, code review, and log and fault analysis \cite{apostolou2026agenticaiindustryadoption}. This shift, however, requires participants to fundamentally change how they approach problems in the future, as agents can increase productivity, but the solutions they generate might be harder to evaluate \cite{10992583}. 

The next step is what is called SE 5.0 \cite{11319541}. Here, the development moves from human-in-the-loop to human-on-loop, where the developers are passively informed of the progress rather than being the driving factor. In other words, human developers will no longer drive development, but will instead be able to gain insight into the process and intervene if needed. The development teams will consist of agents with general domain autonomy with limited need for human oversight. To reach this state, the organization needs to become an AI-driven organization with an AI-first perspective, meaning all processes should be redesigned from an AI-agentic perspective \cite{TowardsAIOrganizations}.

Each of these stages comes with its own challenges that need to be considered. Both from a technological and an organizational perspective. Each transition from one phase to another requires changes to the company's processes and workflow, and requires planning. All related papers presented here provide insights into how some companies push towards advancing into the next phase in SE, however, they raise the question of what organizations need to do to reach this next phase.

\section{Methodology}

This study followed a mixed method to collect and analyze essential data for developing a roadmap towards becoming an AI-first organization. We organized a workshop with an embedded systems company and used its output for structuring this study. The detailed process for data collection and analysis is presented in the subsequent sections.

\subsection{Data Collection}

The workshop was carried out in April 2026 for 2 hours and 30 minutes at the company's premises. The company itself is a large embedded software company with approximately 20,000 employees worldwide. The workshop consisted of 40 participants across different roles, including equal parts Scrum Masters and management, Software Architects, and product owners. Data collection was mainly done in two phases, 

\begin{itemize}

\item Data were collected in the form of answers through a semi-structured Mentimeter-based survey as a part of the workshop. The questions focused on the company's expectations of how AI-based systems can affect their processes and workflows. The questionnaire (20 questions) was intentionally designed to include both objective and subjective question formats. Structured multiple-choice items were used to generate systematic, quantifiable data, whereas open-ended questions allowed participants to articulate their views freely and provide richer qualitative insights. 

 \item Each question-answer sequence was followed by an open-ended brainstorming session that allowed participants not only to elaborate on their responses, but also to raise related issues of importance and share real-world experiences from their practice. These sessions were both recorded 
 and transcribed.

\end{itemize}

\begin{table}[ht]
\centering
\caption{Mentimeter questions used and their type}
\scriptsize
\setlength{\tabcolsep}{1.5pt}
\setlength{\arrayrulewidth}{0.2pt}

\begin{tabular}{L{0.80\textwidth}  L{0.20\textwidth}}
\hline
\textbf{Question}  & \textbf{Type} \\
\hline
If you were to compare generative AI in SE to something, what would it be? & Multiple choice \\
\hline
If your AI assistant (Claude Code) worked on your
project for one week, what would you need to do after? & Ranking\\
\hline
If you encountered a defect in AI-generated code, what
do you do? & Multiple choice \\
\hline
How would you like to see AI in your team? & Multiple choice \\
\hline
Which technology will have the most impact? & Score 0--100 \\
\hline
Which roles will need to adapt the most (re-skilling)? & Score 0--100 \\
\hline
How important is it to build competence in-house for the following? & Score 0--100 \\
\hline
What hinders the adoption of these tools? & Open-ended \\
\hline
If you were to invest in improving AI, what would you improve? & Prioritization \\
\hline
If you were to invest in your organization, what would it be? & Prioritization \\
\hline
What will happen to your team in the future (3--5 years)? & Likert-scale \\
\hline
What will happen to your organization (3--5 years)? & Likert-scale \\
\hline
What will happen to software engineering (3--5 years)? & Likert-scale \\
\hline
\label{tab:questions}
\end{tabular}
\end{table}


\subsection{Data Analysis: Mentimeter Answers}
The Mentimeter questionnaire contained five types of items, i) single-choice questions asked participants to select one option from a set of predefined alternatives, ii) open-ended questions allowed participants to provide free-text responses. iii) voting and iv) prioritization questions asked participants to allocate votes or points across multiple options, resulting in a score for each category and iv) Likert-scale rating questions asked participants to rate several items on an ordered scale, for example, from “much less of” (-2) to “much more of” (+2). For the last type, Mentimeter reported both the distribution of responses across scale points and a weighted average for each item. Table \ref{tab:questions} provides a summarized list of questions. An extended version is available with the possible response options for each question in the related repository\footnote{https://github.com/viktorkjellberg/Developing\_a\_Roadmap\_to\_an\_AI-first\_Organization}.

The analysis was adapted to each response type. Single-choice questions were summarized using vote counts. Voting questions were interpreted through the relative scores assigned to each option. Likert-scale questions were analyzed using weighted averages and the distribution of responses across the scale. Open-ended answers were grouped into descriptive categories based on semantic similarity. These descriptive summaries were used to identify the main priorities, expectations, and concerns expressed by participants during the workshop.

The Mentimeter analysis therefore showed what participants prioritized at the moment of answering, while the discussion narratives provided the explanatory layer needed to understand why these priorities mattered in the context of embedded software engineering.

\subsection{Data Analysis: Open-ended discussion}
The analysis of the qualitative data collected through the open-ended discussion was conducted through an inductive thematic analysis to develop a conceptual model from the given answers. It followed the inductive coding approach as described by  Naeem et al.\cite{thematicanalysis}.     

Step \circled{1} consisted of familiarization and transcription of the data.

Step \circled{2}, the first and second authors independently reviewed the transcript material and identified statement-keyword mappings that captured analytically meaningful ideas in the data. Since the coders sometimes selected partly overlapping but not identical transcript segments, and many differences reflected semantic variation rather than substantive disagreement, agreement was resolved through post-coding discussion and consensus refinement. During this process, the coders jointly reviewed the candidate mappings, discussed differences in interpretation, and decided which mappings were sufficiently meaningful and representative to retain for further analysis. After discussion, 99 candidate mappings were reviewed, of which 85 were retained and 14 were discarded. This corresponds to a post-discussion retention rate of 85.86\%.

Step \circled{3}  Each keyword was then given a code that captured the overall core message.

In step \circled{4}, the focus was on identifying patterns and relationships among the codes and commonalities between them. Through this process, it is possible to group the codes into themes and add a layer of abstraction. These abstractions become the common themes found in the discussion and therefore represent the focus of the participants' view in the discussion part of the data collection.

Finally, in step \circled{5}, these themes were explored to gain knowledge into the participants' understanding and opinion on their transition towards an AI-first organization. From the themes, six findings were extracted, which were used to create a roadmap for the organization.

\section{Findings}

\begin{figure}[t]
     \centering
     \includegraphics[width=0.75\linewidth]{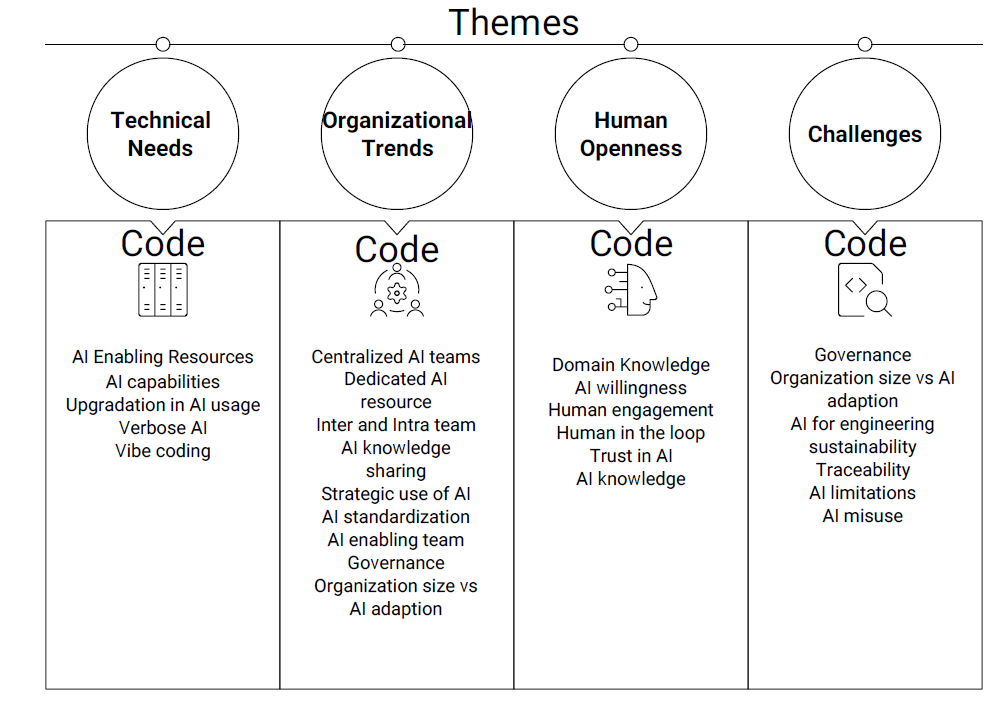}
     \caption{Mapping: Themes to Codes}
     \label{fig:themetocode}
 \end{figure}

The analysis of the answers given through the workshops, along with the discussion, showed that the participants envision significant changes in processes and workflows. The thematic analysis showed four themes that reoccurred throughout the discussion, seen in Figure \ref{fig:themetocode}.

\subsection{RQ1.1: Which technical, organizational, and human factors are expected to be affected by the transition?}

Among respondents to this question, 44\% reported that generative AI would have a significant impact on the industry and that the technology is comparable to the introduction of the World Wide Web (WWW). Additionally, the thematic analysis of the discussion, supported by the Mentimeter questions, identified three themes relevant to answering this question.

\subsubsection{Organizational Trends}

Starting with the \textbf{team composition}, participants believed that typical software engineering roles would decline over the next three to five years, creating room for more AI agents. Participants did not describe this as a replacement for developers, but as a redistribution of work, with agents handling implementation while human engineers focus more on validating outputs, coordinating with stakeholders, and maintaining domain understanding. This leads to the first finding: \textbf{Finding 1: Practitioners perceive the transition towards agentic AI as an organizational transition}. One participant summarized this organizational shift as follows:

\begin{quote}
    \textit{"[...] we have to rethink how we organize. Because AI agents that work autonomously cannot just come into one of our teams, we have to have an AI-first organization."}
\end{quote}

The organization of AI capabilities was identified as a key concern. Some participants argued for the need for centralized support and strategy, through defining common infrastructure, and defining standards. This could help avoid fragmented adoption and ensure that workflows follow shared guardrails, as explained by one participant:

\begin{quote}
    \textit{"You might also wanna consider [...] having an enabling team [...] that goes out to help specific teams for a period of time. Then you also have this continuity of guardrails and governance, and so on, by the enabling team going out and assisting where needed."}
\end{quote}

However, participants also identified the risk of an overly high concentration of capabilities in a centralized manner. Specialized and centralized teams might become detached from everyday engineering work and create dependency on a small group of experts:

\begin{quote}
    \textit{"This [centralized] team becomes a kind of super team for AI. They know a lot, and then the entire organization is left behind [...]."}
\end{quote}

To limit this risk, some participants argued for distributing the knowledge of implementing AI agents to each team instead. Instead of a centralized AI team, each team would be responsible for implementing its own agents to meet its needs. Each team would have resources full-time, or a certain percentage dedicated to AI agent development, and be the team's AI ambassador. However, participants also noted that partial allocation is difficult to sustain when project deadlines compete for attention:

\begin{quote}
    \textit{"These hybrid teams are good in theory, but they are super hard to manage in a good way when you have a project with urgent things and so on."}
\end{quote}

Regardless of how much time engineers should dedicate to AI development and integration, the need for knowledge sharing between team members and teams was identified as an enabling factor. A dedicated community for AI emerged as a potential solution, where engineers could meet and work together towards building a common platform and frameworks. 

\begin{quote}
    \textit{"So, the thing I've seen work best is to create cross-cutting communities of practice or people that have dedicated time in different teams to meet and develop a platform together."}
\end{quote}

Overall, participants favored an organizational model that combines centralized standards and governance with distributed implementation capacity. Cross-team communities of practice and AI ambassadors were seen as important mechanisms for sharing knowledge, avoiding duplication, and aligning local experiments with organizational strategy. This leads to the second finding: \textbf{Finding 2: Participants favored a federated AI community with centralized governance and distributed implementation through AI ambassadors}.

\subsubsection{Technical Needs} related to developing and experimenting with the integration of an agent were identified and emerged as the second theme. The participants recognized the need for technical enablers to advance toward the next step towards becoming an AI-first organization. One need was an experimental environment where developers could test agents without affecting the development environment. The implementation of agentic processes also requires the appropriate hardware resources.

\begin{quote}
    \textit{"So maybe we need to sync to build a sandbox or something which is isolated from the company's internal network to do all the experiments we want."}
\end{quote}

Another enabling factor mentioned was the AI agents' own capability. The ability to use these agents are to a large extent dependent on the capabilities of the LLMs used. To use these agents successfully, participants expressed a need for competence about their abilities and the maturity of these agents. 
This leads to the third finding: \textbf{Finding 3: AI adoption requires safe experimentation and embedded-aware technical infrastructure.}

\subsubsection{Human Openness and Engagement toward AI} relates to the practitioner's attitude towards integrating AI into their systems and how they think that the human factor would be affected by the transition. The participants were asked how they would like to see AI agents implemented in their teams, and a majority (70 \%) answered that they would like to see developers who use AI agents, instead of as a replacement. 
In other words, keeping humans in the loop was expressed as an important factor in the future of AI adaptation, and expressed by one participant as:

\begin{quote}
    \textit{"I mean, you want to be assisted by AI, not replaced by assistants."}
\end{quote}

With that said, many participants believed that their roles in software engineering would change in the coming years and that many current engineers would need to adapt through retraining. Participants predicted that programmers were the role most at risk of needing retraining. As a follow-up question, they were asked which competencies should be focused on to build in the company, in which developing agents, using AI for SE, and prompt engineering were identified as the most needed skills. However, domain knowledge was highlighted as the most valued type of knowledge employees have. The participants expressed that when the need for programmers and other software engineering roles decreases, the value of employees lies in their domain knowledge. 

Finally, AI adoption was described not only as a technical change but also as a competence-building process, and can be summarized as follows. \textbf{Finding 4: Human expertise shifts toward domain knowledge, supervision, validation, and accountability.}

\subsection {RQ1.2: What challenges should be considered when developing the roadmap for such a transition?}

Throughout the discussion, the participants identified a few challenges related to the transformation toward becoming an AI-first organization.
When asked what the main hindrance to the adoption of these tools was, the participants expressed that a lack of strategy was a major hindrance and that there was a need for clear guidance and a common way of working. The next concern identified was the lack of adequate knowledge to use AI tools efficiently.

From the following discussions, we identified our fourth theme, \emph{Challenges in AI adaptation}. One of the identified challenges was \emph{Governance}. One of the participants stated,

\begin{quote}
    \textit{"We are only allowed to use the tools on an approved list. [...] We had a lot of discussion on how to get approved to use MCP."}
\end{quote}

Yet another concern was \emph{loosing governance of their own code \& design}. As AI is being heavily used for programming and even designing the participants feared the loosing the understandability of their own code could be a problem in near future. 

Another issue that came up as a part of the discussion was, that AI adoption might create an illusion of productivity. Participants warned by stating \emph{"AI for software engineering, not to just jump on this because AI gives a illusion of productivity"}. In embedded systems and software product-line contexts, short-term gains might be offset by long-term costs in maintainability, traceability, architectural integrity, and technical debt. Therefore, AI adoption should be evaluated through the broader lens of \emph{AI for engineering sustainability}, where the central question is not only whether AI accelerates development but also whether it helps keep software products and product lines sustainable over time. 

Participants identified several \emph{limitations of current AI tools} when applied to embedded and industrial software engineering. One concern was that AI increases the amount of generated code. As they stated \emph{"We are producing much more code and need much more tests, reviews, and validations"}. Participants also noted that tool performance is uneven across programming languages. While AI coding assistants might perform well for mainstream languages such as Python and Java, they are less reliable for C, Assembly, Ada, Haskell, COBOL, and other languages that remain important in the embedded software domain. Finally, participants questioned whether AI-generated test suites and high coverage metrics provide meaningful assurance, since 100\% coverage might simply indicate that code has been executed rather than that the system has been properly validated. These issues position AI limitations as a central challenge for AI-first embedded software engineering, especially where long-term sustainability, safety, and industrial-grade verification are required. This leads to \textbf{Finding 5: Participants identified a productivity-validation tension, in which local AI acceleration might increase downstream costs}.

Participants identified AI misuse as a challenge when AI agents optimize for superficial task completion rather than the underlying engineering intent. As a participant stated, the agent began modifying test cases rather than fixing the actual code, thereby weakening the tests' role as independent verification artifacts.

\begin{quote}
    \textit{"We had a quite interesting issue in one of the teams where the Agent actually started fixing test cases instead of solving the code, and I think it was actually found because we still have sharp knives in the development team."}
\end{quote}

Participants also noted that AI-generated solutions might overlook embedded-system constraints, such as whether the resulting software fits on the target micro-controller. In addition, excessive code generation can shift the burden to review and validation. Rather than saving time, large volumes of AI-generated code might require substantial human effort to inspect and understand. These concerns suggest that \emph{AI misuse} in embedded software engineering includes not only incorrect outputs but also misaligned task execution, erosion of test integrity, neglect of hardware constraints, and hidden downstream costs. \textbf{Finding 6: AI misuse can lead to misaligned task execution through ignored constraints and erosion of test integrity}.
Finally, as an overall outcome of this workshop, the participants identified AI team formation as a key organizational challenge. A centralized AI team might provide expertise, governance, and standardization, but it can also become a bottleneck, leaving the wider organization dependent on a small group of specialists. Conversely, fully distributed AI development might lead to fragmented practices, duplicated effort, and inconsistent agent frameworks. Participants therefore favored a hybrid or federated model as discussed in Section 4.1.

\subsection{RQ1: How do embedded software development engineers envision the transition to an AI-first organization?}
\begin{figure}[htbp!]

    \includegraphics[width=0.90\textwidth]{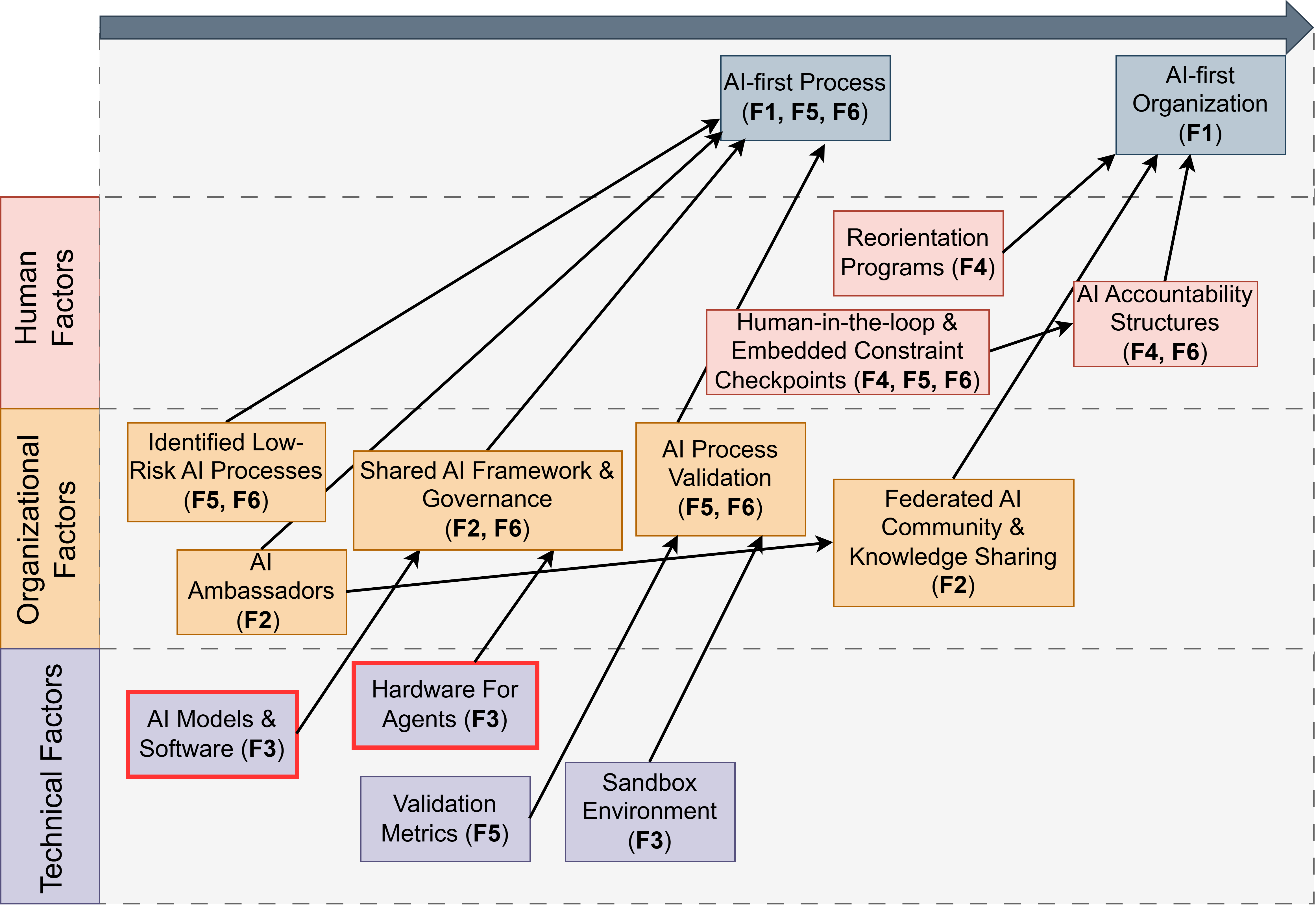}
    \centering
  \caption{Roadmap for transitioning to an AI-first organization. The x-axis represents time. Boxes with a red square around are push factors, and AI-first Process and Organization are both pull factors. All related Findings connected to each stage are indicated by the Finding (F) and the number.}
  \label{fig:rmap}
\end{figure}

The roadmap, as depicted in Figure \ref{fig:rmap}, translates the empirical findings from the thematic analysis into three coordinated dimensions: technical, human, and organizational. Rather than treating the transformation as a purely technical challenge, the roadmap reflects the participants' view that the transition requires changes across technical, organizational design, and human roles. Each element in the roadmap is grounded in one or more findings from the analysis of the transcribed workshop. The order and dependencies between actions in the roadmap were based on results from the open-ended discussion and speculative deliberation among the authors when needed.

The overall goal of the roadmap is to become an AI-first organization, meaning that organizational processes should adopt an AI-first perspective and be optimized for implementing AI agents. This target is grounded in Finding 1, which shows that participants view the transformation as affecting the whole organization. They emphasize that applying the agent directly to the current processes is not possible; instead, the entire process needs to be redesigned. However, to achieve this, there is an intermediate step in the roadmap, \textbf{AI-first Process}, which focuses on redesigning one or a few processes in this manner. This should be seen as the first concrete step towards becoming an AI-first organization.

The technical layer of the roadmap outlines the infrastructure required for the transformation. \textbf{AI Models \& Software} and \textbf{Hardware For Agents} are both enabling factors in the sense that they are the basis for the possibility of the whole transformation. AI Models \& Software indicate that the transformation depends on the capabilities of the underlying LLMs, which are the basis for AI agents, as emphasized by the participants. Their capabilities make the transformation possible but do not, on their own, guarantee its success. The same can be said for the Hardware for Agents, which indicates the need for establishing sufficient, available hardware to integrate and run these agents. Based on the hardware and models, a \textbf{Shared AI Framework \& Governance} can be established, as supported by Findings 2 and 6. This is to ensure a unified approach to working with these agents across the organization. 

The participants identified a need for well-defined \textbf{Validation Metrics} to ensure that implementations actually achieve the desired results, as supported by Findings 5 and 6. The organization, therefore, needs to decide what defines a successful implementation and how to measure it. These metrics need to have a broader scope than just the local effect on productivity, and they should measure the effect on both preceding and subsequent processes in the development chain. This is to ensure that the implementation increases local productivity while not negatively affecting other parts of the development chain. 

In addition to Validation Metrics, there is a need to establish a sufficient \textbf{Sandbox Environment} where developers can test their agent process implementations before deploying them to the real development environment. This is supported by Finding 3, which indicates the usefulness of such an environment. This environment should include the necessary restrictions and constraints of the embedded software domain in which the organization acts. Validation Metrics and the Sandbox Environment both contribute to a unified \textbf{AI Process Validation}, which should be the organizational process governing how and when to measure effects and to unify measurements across the organization. 

The organization should start with \textbf{Identified Low-Risk AI Processes} at the beginning of the transformation. These can be processes that are more or less independent of the rest of the development chain, with a low risk of increasing downstream costs, as supported by Finding 5. These initial processes should also be non-essential, in which the organization can afford to make implementation mistakes (Finding 6). 

Once at least one such low-risk process has been identified, the team responsible for it should assign an \textbf{AI Ambassador}. One person is responsible for implementing the agents into the process, as supported by Finding 2. When all the steps discussed until now have been implemented, the organization is ready for its first redesigned AI-first process. 

When the organization is ready to go from an AI-first Process to an AI-first organization, there is a greater focus on the identified human factors. The First is to identify the \textbf{Human-in-the-loop \& Embedded Constraints Checkpoints} based on activities that must be handled by employees, either legally or technically. These can serve as checkpoints for validating the generated products that cannot be reliably automated, as supported by Finding 4. The change in the developer's role also means that an \textbf{AI Accountability Structure} is needed. In other words, an accountability structure designed for the increased rate of content generation within the development structure. This is supported by both Finding 4 and Finding 6. This includes clarifying who is responsible for the code generated by the agents and ensuring that developers are comfortable using them, even when the quality of the results cannot be guaranteed at 100\%.

In addition to these, there should be a dedicated \textbf{Reorientation Program} for software engineers whose roles will change. According to Finding 4, there should be opportunities for employees to retrain for roles needed in the transformation toward an AI-first organization. This could focus on supervision, validation, and expertise in the specific domain they work in. 

All these steps were identified through analysis of the transcribed workshop and could help the organization transform into an AI-first organization. Many of the steps in the roadmap are not static instances, and their relative positions indicate when the first iteration of each step should occur. Almost all of them will require continuous refinement throughout the transformation and afterward to ensure they are up to date.

\section{Limitation}
The roadmap was developed from a prospective viewpoint of the organization's vision to become an AI-first organization. This means we can't validate its applicability until the transformation has occurred. Therefore, the proposed roadmap remains conceptual until the transformation begins.

\section{Discussion}

The overarching goal of moving towards becoming an AI-first organization is to take advantage of the potential gains in productivity shown when AI-supported tools and AI agents are used in software engineering organizations \cite{cui2026effects,AgentsForUsers,Hitl}. However, to achieve this gained productivity, the whole organization needs to adapt \cite{TowardsAIOrganizations}. Our proposed roadmap therefore includes actions across three separate factors.

Regarding the \textbf{Technical Factors}, our recommendation to establish a sandbox environment aligns with prior work on LLM-based development in embedded systems to ensure a safe environment for developers to test their implementations \cite{11344427}. In the context of an embedded software system, such environments are particularly important because agentic AI might interact not only with source code but also with sensitive build systems. Alongside the sandbox environment, there will also be an increase in computational resources needed. The implemented agents can be costly due to the added need for both CPU and GPU resources \cite{10822885}. Without a scalable infrastructure, agentic workflows might remain local experiments and fragmented across the organization.

The \textbf{Human Factors} identified in the roadmap point to a second important tension that agentic AI might increase development capacity but also change engineers' responsibilities. Prior studies highlight the need for targeted training as software engineering roles evolve and AI-based tools become part of everyday work \cite{10.1145/3652154}. Our findings extend this argument by showing that in embedded software organizations, competence development should focus on engineers' ability to evaluate AI-generated outputs in relation to domain constraints and verification. This is especially important because current agentic systems still require human oversight \cite{apostolou2026agenticaiindustryadoption}. As AI-generated code and content becomes more common, responsibility for the resulting artifact might become unclear. Previous work emphasizes that developers might feel less accountable for AI-generated content \cite{10.1145/3613905.3650764}. The roadmap therefore focuses on explicit human-in-the-loop checkpoints and clarified accountability structures.

When it comes to \textbf{Organizational Factors}, the roadmap highlights the need for a centralized strategy to define shared frameworks and common metrics. However, a purely centralized AI function might become detached and separated from the rest of the organization. On the other hand, a fully decentralized process might lead to duplicated effort and errors in the intersections between teams. The recommended federated structures address this by combining shared standards with team-level implementation through AI ambassadors, which brings capabilities closer to practical implementations.

Several participants warned against interpreting local productivity gains as organizational value. This concern is important in embedded systems, where faster code generation can increase the workload, requiring more reviews or testing further down the production chain. Therefore, the proposed roadmap recommends unified metrics that evaluate impact across the whole development cycle. This perspective aligns with previous studies about the effect of AI agents in software engineering \cite{10.1145/3796563.3796601,wang2026doesagentdevelopmentreflect}.

The roadmap also supports a gradual and risk-aware adoption of agents. The processes should, as a start, be low-risk activities, such as documentation, pull request support, or failure analysis \cite{10.1145/3796563.3796601,sergeyuk_using_2025,10.1145/3652154}. The organization should also consider activities with a high manual workload, which could benefit from automation. As the organization matures, more and more complex activities can be considered, and the focus should shift toward focusing on activities with high potential return. However, each introduced process needs to be fully redesigned from an AI-first perspective to ensure optimal agent performance \cite{TowardsAIOrganizations}. Therefore, starting with the low-risk activities ensures the validity of the new processes and allows the organization to be more experimental with the development of new workflows.

To summarize, the proposed roadmap supports a gradual and risk-aware path. Rather than broadly implementing agents, the organization should focus on identifying candidate processes in which the task is bounded and verification is feasible. The adaptation should focus on knowledge sharing between teams and reorient employees to positions that are likely to be needed for an efficient and productive adaptation of AI agents.

Our proposed roadmap is based on the findings from a single organization's vision toward becoming an AI-first organization. Therefore, the proposed processes and actions reflect this organization's needs, context, and current progress. The overall themes should, however, be generalizable, but the order and dependencies between factors should be reevaluated if applied to other organizations. 


\section{Threats to Validity}

\textbf{Construct validity} might be affected by the design of the Mentimeter questions. The questions used were designed to help the discussion throughout the workshop. The alternatives for each question might be biased toward our views and assumptions as authors, which could affect the participants' answers. This could lead to the participants making decisions based on the narrative of the question formulation. However, combining the Mentimeter question with the follow-up discussion allowed participants to elaborate on their answers.

\textbf{Internal validity} might be affected by the design of the workshop. Participants' views and opinions might have been influenced by the framing or by other participants' opinions. Open discussions might also be dominated by a few passionate voices. However, to mitigate this, the data were collected through both anonymized individual Mentimeter responses and the group discussion. 

\textbf{External validity} is limited by the fact that the study only includes a single case. The findings reflect one large company in the embedded software domain and might not generalize to companies of other sizes or in other domains. The proposed roadmap should therefore be seen as an empirically grounded proposal for similar embedded software organizations undergoing the same transition rather than a general model.

\section{Conclusion and Future Work}

This paper presents a case study of how practitioners in a large embedded software company envision their transition from SE 3.0 (AI-assisted) towards SE 4.0 (Agentic systems). Our findings, based on data collected through a semi-structured workshop with 100 participants, indicate that participants envision the transition will broadly affect the organization by changing team composition, changing role responsibilities, and shifting the required skills of software engineers. The participants envision the transition being supported by a company-wide federated AI community with centralized standards and frameworks for AI implementation, but with the actual implementation carried out by AI ambassadors from each affected team. 

The thematic analysis of the discussion identified 27 codes grouped into four themes related to Technical Needs, Organizational Trends, Human Openness, and Challenges. From these themes, six findings were identified. These findings were used to develop a roadmap outlining the steps required to integrate AI agents into processes to meet the envisioned changes and overcome the challenges expressed by the participants. The roadmap covers steps for overcoming challenges related to the Organizational, Technical, and Human factors identified. 

Taken together, the transition will require broad investments in both technical solutions related to implementation and experimentation with agentic systems, employee reorientation toward new skills and deeper domain knowledge, and a federated community responsible for supporting implementation through standards, knowledge sharing, and measurements.

However, the proposed roadmap is conceptual and needs validation to ensure its practicality. In the current version, each identified factor is placed based on its relation to the two goals and reflects a single organization's perspective. Therefore, future work should include validating the proposed roadmap through an additional workshop where the participants will be allowed to share their thoughts on the roadmap and where each factor could be broken down into finer processes with signals of when each stage is completed and when it would be time to move to the next stage. Another direction is to test the generalizability of the proposed roadmap across multiple organizations in embedded software development.

\section*{Acknowledgment}

The Swedish Research Council partially funded this study under grant number 2024-04687.

\bibliography{ref}

@article{thematicanalysis,
  title={A step-by-step process of thematic analysis to develop a conceptual model in qualitative research},
  author={Naeem, Muhammad and Ozuem, Wilson and Howell, Kerry and Ranfagni, Silvia},
  journal={International journal of qualitative methods},
  volume={22},
  pages={16094069231205789},
  year={2023},
  publisher={SAGE Publications Sage CA: Los Angeles, CA}
}

@INPROCEEDINGS{10822885,
  author={Chkirbene, Zina and Hamila, Ridha and Gouissem, Ala and Devrim, Unal},
  booktitle={2024 IEEE 21st International Conference on Smart Communities: Improving Quality of Life using AI, Robotics and IoT (HONET)}, 
  title={Large Language Models (LLM) in Industry: A Survey of Applications, Challenges, and Trends}, 
  year={2024},
  volume={},
  number={},
  pages={229-234},
  doi={10.1109/HONET63146.2024.10822885}
}

@article{10.1145/3652154,
author = {Russo, Daniel},
title = {Navigating the Complexity of Generative AI Adoption in Software Engineering},
year = {2024},
issue_date = {June 2024},
publisher = {Association for Computing Machinery},
address = {New York, NY, USA},
volume = {33},
number = {5},
issn = {1049-331X},
doi = {10.1145/3652154},
journal = {ACM Trans. Softw. Eng. Methodol.},
month = jun,
articleno = {135},
numpages = {50}
}

@misc{wang2026doesagentdevelopmentreflect,
      title={How Well Does Agent Development Reflect Real-World Work?}, 
      author={Zora Zhiruo Wang and Sanidhya Vijayvargiya and Aspen Chen and Hanmo Zhang and Venu Arvind Arangarajan and Jett Chen and Valerie Chen and Diyi Yang and Daniel Fried and Graham Neubig},
      year={2026},
      eprint={2603.01203},
      archivePrefix={arXiv},
      primaryClass={cs.AI},
      url={https://arxiv.org/abs/2603.01203}, 
}

@inproceedings{10.1145/3613905.3650764,
author = {Englhardt, Zachary and Li, Richard and Nissanka, Dilini and Zhang, Zhihan and Narayanswamy, Girish and Breda, Joseph and Liu, Xin and Patel, Shwetak and Iyer, Vikram},
title = {Exploring and Characterizing Large Language Models for Embedded System Development and Debugging},
year = {2024},
isbn = {9798400703317},
publisher = {Association for Computing Machinery},
address = {New York, NY, USA},
url = {https://doi.org/10.1145/3613905.3650764},
booktitle = {Extended Abstracts of the CHI Conference on Human Factors in Computing Systems},
articleno = {150},
numpages = {9},
location = {Honolulu, HI, USA},
series = {CHI EA '24}
}

@INPROCEEDINGS{11344427,
  author={Abtahi, Seyed Moein and Azim, Akramul},
  booktitle={2025 IEEE International Conference on Collaborative Advances in Software and COmputiNg (CASCON)}, 
  title={Securing LLM-Generated Embedded Firmware Through AI Agent-Driven Validation and Patching}, 
  year={2025},
  volume={},
  number={},
  pages={210-218},
  doi={10.1109/CASCON66301.2025.00044}}

@inproceedings{NIPS2017_3f5ee243,
 author = {Vaswani, Ashish and Shazeer, Noam and Parmar, Niki and Uszkoreit, Jakob and Jones, Llion and Gomez, Aidan N and Kaiser, \L ukasz and Polosukhin, Illia},
 booktitle = {Advances in Neural Information Processing Systems},
 editor = {I. Guyon and U. Von Luxburg and S. Bengio and H. Wallach and R. Fergus and S. Vishwanathan and R. Garnett},
 pages = {},
 publisher = {Curran Associates, Inc.},
 title = {Attention is All you Need},
 volume = {30},
 year = {2017}
}

@ARTICLE{11319541,
  author={Jiang, Zhen Ming and Hassan, Ahmed E. and Zimmermann, Thomas and Harman, Mark},
  journal={IEEE Software}, 
  title={AIware in the Foundation Model Era}, 
  year={2026},
  volume={43},
  number={1},
  pages={26-30},
  doi={10.1109/MS.2025.3624446}}

@inproceedings{10.1145/3796563.3796601,
author = {Varanasi, Dhiraj SM and Reddy, Y. Raghu and Potdar, Sandip and Warudkar, Himanshu S},
title = {Adoption of AI Assisted Coding – Cognitive Barriers, Motivational Factors and Productivity Gains},
year = {2026},
isbn = {9798400725036},
publisher = {Association for Computing Machinery},
address = {New York, NY, USA},
doi = {10.1145/3796563.3796601},
booktitle = {Proceedings of the 19th Innovations in Software Engineering Conference},
articleno = {6},
numpages = {11},
location = {
},
series = {ISEC '26}
}

@INPROCEEDINGS{sun2026agenticpipelinesembeddedsoftware,
  author={Sun, Simin and Staron, Miroslaw},
  booktitle={2026 IEEE International Conference on Software Analysis, Evolution and Reengineering (SANER)}, 
  title={Agentic Pipelines in Embedded Software Engineering: Emerging Practices and Challenges}, 
  year={2026},
  volume={},
  number={},
  pages={136-146},
  doi={10.1109/SANER67736.2026.00024}}

@misc{ferino2025walkingtightropellmssoftware,
      title={Walking the Tightrope of LLMs for Software Development: A Practitioners' Perspective}, 
      author={Samuel Ferino and Rashina Hoda and John Grundy and Christoph Treude},
      year={2025},
      eprint={2511.06428},
      archivePrefix={arXiv},
      primaryClass={cs.SE},
      url={https://arxiv.org/abs/2511.06428}, 
}

@INPROCEEDINGS{10992583,
  author={Parnin, Chris and Soares, Gustavo and Pandita, Rahul and Gulwani, Sumit and Rich, Jessica and Henley, Austin Z.},
  booktitle={2025 IEEE International Conference on Software Analysis, Evolution and Reengineering (SANER)}, 
  title={Building Your Own Product Copilot: Challenges, Opportunities, and Needs}, 
  year={2025},
  volume={},
  number={},
  pages={338-348},
  doi={10.1109/SANER64311.2025.00039}}

@inproceedings{10.1145/3643795.3648377,
author = {Ramler, Rudolf and Moser, Michael and Fischer, Lukas and Nissl, Markus and Heinzl, Rene},
title = {Industrial Experience Report on AI-Assisted Coding in Professional Software Development},
year = {2024},
isbn = {9798400705793},
publisher = {Association for Computing Machinery},
address = {New York, NY, USA},
doi = {10.1145/3643795.3648377},
booktitle = {Proceedings of the 1st International Workshop on Large Language Models for Code},
pages = {1–7},
numpages = {7},
location = {Lisbon, Portugal},
series = {LLM4Code '24}
}

@article{sergeyuk_using_2025,
title = {Using AI-based coding assistants in practice: State of affairs, perceptions, and ways forward},
journal = {Information and Software Technology},
volume = {178},
pages = {107610},
year = {2025},
issn = {0950-5849},
doi = {https://doi.org/10.1016/j.infsof.2024.107610},
url = {https://www.sciencedirect.com/science/article/pii/S0950584924002155},
author = {Agnia Sergeyuk and Yaroslav Golubev and Timofey Bryksin and Iftekhar Ahmed}
}

@InProceedings{TowardsAIOrganizations,
author="Bosch, Jan
and Olsson, Helena Holmstr{\"o}m",
editor="Taibi, Davide
and Smite, Darja",
title="Towards AI-Driven Organizations",
booktitle="Software Engineering and Advanced Applications",
year="2026",
publisher="Springer Nature Switzerland",
address="Cham",
pages="280--295",
isbn="978-3-032-04207-1"
}

@misc{apostolou2026agenticaiindustryadoption,
      title={Agentic AI in Industry: Adoption Level and Deployment Barriers}, 
      author={Spyridon Alvanakis Apostolou and Jan Bosch and Helena Holmström Olsson},
      year={2026},
      eprint={2605.14675},
      archivePrefix={arXiv},
      primaryClass={cs.SE},
      url={https://arxiv.org/abs/2605.14675}, 
}

@article{cui2026effects,
  title={The effects of generative AI on high-skilled work: Evidence from three field experiments with software developers},
  author={Cui, Kevin Zheyuan and Demirer, Mert and Jaffe, Sonia and Musolff, Leon and Peng, Sida and Salz, Tobias},
  journal={Management Science},
  year={2026},
  publisher={INFORMS}
}

@INPROCEEDINGS{AgentsForUsers,
  author={Feng, Sidong and Du, Changhao and Liu, Huaxiao and Wang, Qingnan and Lv, Zhengwei and Huo, Gang and Yang, Xu and Chen, Chunyang},
  booktitle={2025 IEEE/ACM 47th International Conference on Software Engineering: Software Engineering in Practice (ICSE-SEIP)}, 
  title={Agent for User: Testing Multi - User Interactive Features in TikTok}, 
  year={2025},
  volume={},
  number={},
  pages={57-68},
  doi={10.1109/ICSE-SEIP66354.2025.00011}}

@INPROCEEDINGS{Hitl,
  author={Takerngsaksiri, Wannita and Pasuksmit, Jirat and Thongtanunam, Patanamon and Tantithamthavorn, Chakkrit and Zhang, Ruixiong and Jiang, Fan and Li, Jing and Cook, Evan and Chen, Kun and Wu, Ming},
  booktitle={2025 IEEE/ACM 47th International Conference on Software Engineering: Software Engineering in Practice (ICSE-SEIP)}, 
  title={Human-In-The-Loop Software Development Agents}, 
  year={2025},
  volume={},
  number={},
  pages={342-352},
  doi={10.1109/ICSE-SEIP66354.2025.00036}}

\end{document}